\documentclass[12pt,thmsa]{article}
\usepackage{amsfonts}

\input tcilatex
\renewcommand\baselinestretch{1.0}
\begin{document}

\title{NON - BOLTZMANN EQUILIBRIUM PROBABILITY DENSITIES FOR NON - LINEAR L\'{E}VY
OSCILLATOR}
\author{$A.\,Chechkin^{a},V.\,Gonchar^{a},$ \and $R.\,Gorenflo^{b}$, $%
\,F.\,Mainardi^{c},\,L.\,Tanatarov^{a}$ \\
(a) Kharkov Institute for Theoretical Physics\\
National Science Center\\
``Kharkov Institute of Physics and Technology'',\\
Akademicheskaya st.1, Kharkov 310108, Ukraine,\\
(b)Department of Mathematics and Computer Science\\
Free University of Berlin\\
Arnimallee 3, D - 14195 Berlin,\\
Germany \\
(c)$\,$Department of Physics, University of Bologna, Via Irnerio 46,\\
I - 40126 Bologna, Italy}
\maketitle

\begin{abstract}
\renewcommand\baselinestretch{1.0} We study, both analytically and by
numerical modeling the equilibrium probability density function for an non -
linear L\'{e}vy oscillator with the L\'{e}vy index $\alpha $, $1\leq \alpha
\leq 2$, and the potential energy $x^{4}$. In particular, we show that the
equilibrium PDF is bimodal and has power law asymptotics with the exponent $%
-(\alpha +3)$.
\end{abstract}

\strut \textit{PACS}: 05.10 Gg, 05.40 Fb

\smallskip


\section{Starting equations}

Recently, kinetic equations with fractional derivatives have attracted
attention as a possible tool for the description of anomalous diffusion and
relaxation phenomena, see, e.g., recent review \cite{Klafter}, semi - review
papers \cite{Zaslavsky},\cite{Saichev},\cite{Mainardi} and references on
earlier studies therein. It was also recognized \cite{Compte}, \cite{Scalas}
that the fractional kinetic equations may be viewed as ``hydrodynamic''
(that is, long - time and long - space) limits of the CTRW (Continuous Time
Random Walk) scheme \cite{Montroll}, which was successfully applied to the
description of anomalous diffusion phenomena in many areas, e.g., turbulence 
\cite{Klafter1}, disordered medium \cite{Bouchaud}, intermittent chaotic
systems \cite{Zumofen}, etc. However, the kinetic equations have two
advantages over a random walk approach: firstly, they allows one to explore
various boundary conditions (e.g., reflecting and/or absorbing) and,
secondly, to study diffusion and/or relaxation phenomena in external fields,
both possibilities are difficult to realize in the framework of CTRW (we
point, however, to the paper \cite{Barkai}, in which a fractional kinetic
equation was obtained from generalized CTRW). Fractional kinetic equations
can be divided into three classes: the first one, describing Markovian
processes, contains equations with fractional space or velocity derivatives
and the first time derivative, the second one, describing non - Markovian
processes, contains equations with fractional time derivative, and the third
class, naturally, contains both fractional space and time derivatives, as
well. In this paper we deal with a one - dimensional kinetic equation
belonging to the first class, namely, with the Fractional Symmetric Einstein
- Smoluchowski Equation (FSESE), which, from one hand, is a natural
generalization of the diffusion - like equation with the symmetric
fractional space derivative \cite{Saichev},\cite{Paradisi} and, from the
other hand, is a Markovian generalization of the Einstein - Smoluchowski
kinetic equation, which describes a motion of a particle subjected to a
white Gaussian noise in a strong friction limit, see, e.g., \cite
{Chandrasekhar}. From this point of view, the FSESE describes a motion of a
particle subjected to a white L\'{e}vy noise, also in a strong friction
limit \cite{Chechkin}.

In dimensionless units the one - dimensional FSESE has the form 
\begin{equation}
\frac{\partial f}{\partial t}=-\frac{\partial }{\partial x}\left( Ff\right) +%
\frac{\partial ^{\alpha }f}{\partial \left| x\right| ^{\alpha }}%
\,,\,t>0,\,x\in \Bbb{R},\quad \,f(x,0)=\delta (x)\,,  \tag{1.1}
\end{equation}
where $f(x,t)$ is the probability density function, $F$ is the deterministic
external force, $\alpha $ is the L\'{e}vy index, $0<\alpha \leq 2$, $%
\partial ^{\alpha }/\partial \left| x\right| ^{\alpha }$ is the symmetric
fractional space derivative [2, 11], which is defined, for a ``sufficiently
well - behaved function $\phi (x)$`` through its Fourier - transform $\hat{%
\phi}(k)$ as 
\begin{equation}
\frac{d^{\alpha }}{d\left| x\right| ^{\alpha }}\phi (x)\div -\left| k\right|
^{\alpha }\hat{\phi}(k),  \tag{1.2}
\end{equation}
or in terms of the Riemann - Liouville derivatives as 
\begin{equation}
\frac{d^{\alpha }}{d\left| x\right| ^{\alpha }}\phi (x)=-\frac{1}{2\cos (\pi
\alpha /2)}\left[ D_{+}^{\alpha }\phi (x)+D_{-}^{\alpha }\phi (x)\right] , 
\tag{1.3}
\end{equation}
where $\alpha >0,\,\alpha \neq 1,3,...$ , 
\begin{equation}
D_{+}^{\alpha }\phi (x)=\frac{1}{\Gamma (1-\alpha )}\frac{d}{dx}%
\int\limits_{-\infty }^{x}\frac{\phi (x^{\prime })dx^{\prime }}{(x-x^{\prime
})^{\alpha }},\,\,\,\,D_{-}^{\alpha }\phi (x)=-\frac{1}{\Gamma (1-\alpha )}%
\frac{d}{dx}\int\limits_{x}^{\infty }\frac{\phi (x^{\prime })dx^{\prime }}{%
(x^{\prime }-x)^{\alpha }}  \tag{1.4}
\end{equation}
for $0<\alpha <1$. For $\alpha \geq 1$%
\begin{equation}
D_{\pm }^{\alpha }\phi (x)=\frac{(\pm 1)^{n}}{\Gamma (n-\alpha )}\frac{d^{n}%
}{dx^{n}}\int\limits_{0}^{\infty }\xi ^{n-\alpha -1}\phi (x\mp \xi )d\xi , 
\tag{1.5}
\end{equation}
$n-1<\alpha \leq n\in \Bbb{N}$. The derivatives (1.4) and (1.5) are
characterized in their Fourier representation by 
\begin{equation}
D_{\pm }^{\alpha }\phi (x)\div (\mp ik)^{\alpha }\hat{\phi}(k),  \tag{1.6}
\end{equation}
where $(\mp ik)^{\alpha }=\left| k\right| ^{\alpha }\exp \left( \mp \frac{%
i\alpha \pi }{2}sgnk\right) $ . One can easily recover Eq. (1.2) by
combining Eq. (1.3) with Eq. (1.6). A detailed theory of the Riemann -
Liouville and other forms of fractional derivatives is presented in \cite
{Samko}.

The solution of Eq. (1.1) in the force - free case, $F=0$, is known to be a
probability density function (PDF) for an $\alpha -$ stable symmetric
process. The solution of Eq. (1.1) was also obtained for the constant field, 
$F=const$ on the whole axis, $-\infty <x<\infty $, and for a linear L\'{e}vy
oscillator \cite{Fogedby}. As far as the authors know, Eq. (1.1) was not
considered for a more complicated potential fields. In the present paper we
study the equilibrium solution of the FSESE for non - linear L\'{e}vy
oscillator with the potential energy 
\begin{equation}
U(x)=\frac{x^{4}}{4},\,\,\,F=-\frac{dU}{dx}.  \tag{1.7}
\end{equation}
The models with the potential energy (1.7) play an important role in the
theory of dynamical chaos \cite{Bolotin} and in the theory of Brownian
motion in an open auto - oscillation systems \cite{Klimontovich}, and have
various applications \cite{Bulavin}. One may expect that the models of non -
linear L\'{e}vy oscillator will also possess an important place in the
theory of the systems influenced by non - Gaussian L\'{e}vy noises and
obeying fractional kinetic equations. For our purposes we pass to the
equation for the characteristic function $\hat{f}(k)$ , 
\begin{equation}
\hat{f}(k)=\int\limits_{-\infty }^{\infty }dxf(x)\exp (ikx).  \tag{1.8}
\end{equation}
For the equilibrium state $\hat{f}(k)$ obeys the equation, which follows
from Eqs. (1.1) and (1.7): 
\begin{equation}
\frac{d^{3}\hat{f}_{eq}}{dk^{3}}=sgn(k)\left| k\right| ^{\alpha -1}\hat{f}%
_{eq}(k),  \tag{1.9}
\end{equation}
where the index ``eq.'' denotes equilibrium solution. Below in the paper we
omit subscript ``eq'' for brevity. The characteristic function obeys the
following conditions.

1. $\widehat{f}\left( k\right) =\widehat{f^{*}}\left( k\right) =\widehat{f}%
\left( -k\right) ,$where the asterisk implies complex conjugate. The first
equality is a consequence of the Khintchine theorem about reality of the
characteristic function for the symmetric PDF, whereas the second equality
is the consequence of the Bochner - Khintchine theorem about positive
definiteness of the characteristic function. In fact, one can see that the
solution of Eq. (1.9) obeys the condition 1.

2. The boundary conditions for $\widehat{f}(k)$ are as follows: 
\begin{equation}
\widehat{f}(0)=1\,,\quad \widehat{f}(\pm \infty )=0\,,\quad \frac{d\widehat{f%
}(0)}{dk}=0\,.  \tag{1.10}
\end{equation}

\strut The first property stems from normalization of the PDF, $\stackrel{%
\infty }{\stackunder{-\infty }{\int }}f(x)dx=1.$

\strut The second property stems from the existence of the PDF.

\strut As to the third property, we remind that the integer moments of the
PDF (if exist) are connected with the derivatives of the characteristic
function at $k=0$ as 
\[
\left\langle x^{p}\right\rangle =\frac{1}{i^{p}}\frac{d\widehat{f}^{\left(
p\right) }\left( 0\right) }{dk^{p}},\,\,\,p=1,2... 
\]
The third property is the consequence of this theorem: 
\[
\frac{d\widehat{f}^{\left( p\right) }\left( 0\right) }{dk^{p}}%
=0,\,\,\,\,\,p=1,\,3,\,5,\,...\,, 
\]
because the PDF is a symmetric function, and hence, all odd moments are
equal zero. The last equality is valid for those odd $p$, for which the $p$
- th moments of the PDF exist.

To make results more transparent, the rest part of the paper will be
organized as follows. In Sect.2 we recall the results for non - linear
Brownian oscillator, $\alpha =2$. In Sect.3 we get an equilibrium solution
for an non - linear Cauchy oscillator, $\alpha =1$. In Sect.4, which is the
main one, we present analytical and numerical results on equilibrium
solutions for non - linear L\'{e}vy oscillators with the L\'{e}vy indexes
lying between 1 and 2, Brownian and Cauchy oscillators being their
particular cases. Finally, in Sect.5 we present the discussion and summarize
the results. Some subsidiary studies, which support the results obtained in
Sec.4, are presented in Appendix A. An integral which is necessary for
estimating asymptotics of the PDF\ is calculated in Appendix B.

\section{Equilibrium solution for a non - linear Brownian oscillator}

In this Section, which is presented here mainly for the methodical purposes,
we remind the equilibrium solution for non - linear Brownian oscillator, $%
\alpha =2$. It is well - known, that in this case there is no need to pass
to the equation (1.9); on the contrary, the starting point is the stationary
equation, which follows from Eqs. (1.1) and (1.7):

\begin{equation}
\frac{df(x)}{dx}=-\frac{dU}{dx}f(x).  \tag{2.1}
\end{equation}
It has the Boltzmann solution

\begin{equation}
f(x)=C\exp (-U(x)),  \tag{2.2}
\end{equation}
where the constant $C$ is determined from normalization condition. For a non
- linear quartic Brownian oscillator

\begin{equation}
C=\frac{\sqrt{2}}{\Gamma (1/4)}.  \tag{2.3}
\end{equation}

We are also in position to get the characteristic function by making Fourier
transform of equilibrium solution (2.2), (2.3) and, then expanding $exp(ikx)$
into the power series and integrating over $x$ each term separately:

\begin{equation}
\hat{f}(k)=\sum\limits_{j=0}^{\infty }c_{j}k^{2j},  \tag{2.4}
\end{equation}
where

\begin{equation}
c_{j}=\frac{(-1)^{j}2^{j}}{(2j)!}\frac{\Gamma \left( \frac{2j+1}{4}\right) }{%
\Gamma (1/4)}.  \tag{2.5}
\end{equation}
In Sect.. 4 we show that this solution is a particular case of that for a
non - linear L\'{e}vy oscillator.

\section{Equilibrium solution for non - linear Cauchy oscillator}

For this case we start from Eq. (1.9) with $\alpha =1$. The solution is

\begin{equation}
\hat{f}(k)=\frac 2{\sqrt{3}}\exp \left( -\frac{\left| k\right| }2\right)
\cos \left( \frac{\sqrt{3}\left| k\right| }2-\frac \pi 6\right)  \tag{3.1}
\end{equation}
.This solution can be expanded into the power series of $|k|$ as

\begin{equation}
\hat{f}(k)=\sum\limits_{j=0}^\infty c_j\left| k\right| ^j,  \tag{3.2}
\end{equation}

where

\begin{equation}
c_j=\frac 1{2j!}\left[ \left( 1-\frac i{\sqrt{3}}\right) \exp \left( \frac{%
2\pi i}3j\right) +\left( 1+\frac i{\sqrt{3}}\right) \exp \left( \frac{4\pi i}%
3j\right) \right] .  \tag{3.3}
\end{equation}

Making an inverse Fourier transform of Eq. (3.1), we get equilibrium PDF for
the Cauchy oscillator,

\begin{equation}
f(x)=\frac{1}{\pi (1-x^{2}+x^{4})}.  \tag{3.4}
\end{equation}
\strut Clearly, $f(x)\geq 0$ and $\stackrel{\infty }{\stackunder{-\infty }{%
\int }}f(x)dx=1$, so $f$ is a PDF.

Equation (3.4) clearly indicates strongly non - Boltzmann character of the
equilibrium PDF for non - linear Cauchy oscillator. In Fig.1 both PDFs for
Brownian and Cauchy oscillators are shown in a linear (at the top) and in a
semi - logarithmic (at the bottom) scales. The two important distinctive
features of the equilibrium PDF for the Cauchy oscillator are (i) the
bimodality, and (ii) power law asymptotic at $x\rightarrow \pm \infty $. The
latter property is clearly visualized in a linear scale, whereas the former
is better shown in a semi - logarithmic scale. It appears that both features
are inherent not only to a non - linear Cauchy oscillator, but also to
equilibrium PDFs of non - linear L\'{e}vy oscillators with the L\'{e}vy
indexes $\alpha $ , such that $1\leq \alpha <2$. In the next Section we get
the solution for an arbitrary $\alpha \,\,$between 1 and 2 and clearly
demonstrate this fact.

\section{Equilibrium solution for non - linear L\'{e}vy oscillator, $1\leq
\alpha \leq 2$}

We turn to the solution of Eq. (1.9) with the boundary conditions (1.10).
One can convince himself that the solution of Eq. (1.9) with the first and
the third conditions from Eqs. (1.10) being taken into account can be
represented as

\begin{equation}
\hat{f}(k)=\Sigma _{1}+ak^{2}\Sigma _{2},  \tag{4.1}
\end{equation}
where

\begin{equation}
\sum\nolimits_1=1+\sum\limits_{j=1}^\infty a_j\left| k\right| ^{j(\alpha
+2)},  \tag{4.2}
\end{equation}

\begin{equation}
\,\,\,\,\sum\nolimits_{2}=1+\sum\limits_{j=1}^{\infty }b_{j}\left| k\right|
^{j(\alpha +2)}.  \tag{4.3}
\end{equation}
and the coefficients $aj,bj$ are obtained by inserting Eqs. (4.2) and (4.3)
into Eq. (1.9) and equating the terms of the same powers of $k$ in the right
- and left - hand sides:

\begin{equation}
a_jj(\alpha +2)(j\alpha +2j-1)(j\alpha +2j-2)=a_{j-1},  \tag{4.4}
\end{equation}

\begin{equation}
b_jj(\alpha +2)(j\alpha +2j+1)(j\alpha +2j+2)=b_{j-1},  \tag{4.5}
\end{equation}

$j\geq 1,\,\,a_{0}=b_{0}=1.$

Obviously, the $a_{j}$ as well as $b_{j}$ tend to zero extremely fast.

The terms $\Sigma _{1}$ and $k^{2}\Sigma _{2}$ are, in fact, two independent
particular solutions of Eq. (1.9). Since the condition at the infinity from
Eq. (1.10) has not been employed yet, the general solution thus depends on
an arbitrary constant $a$. We define it numerically by demanding

\begin{equation}
\hat{f}(k\rightarrow \infty )\rightarrow 0,  \tag{4.6}
\end{equation}
that is,

\begin{equation}
a=-\lim_{k\rightarrow \infty }\frac{\sum\nolimits_1}{k^2\sum\nolimits_2}. 
\tag{4.7}
\end{equation}

Our numerical simulations show that with $k$ increasing the value of $a$
rapidly reaches the constant value, which, of course, depends on $\alpha $.
Further increase of $k$ allows us to get $a$ with higher accuracy, that is,
with more significant digits. It is also worthwhile to note that the radius $%
R$ of convergence for both power sets in numerator and denominator of Eq.
(4.7) is infinite. This fact can be easily shown with the help of the Cauchy
- Hadamard theorem (see, e.g., \cite{Fikhtengoltz}, p. 300), according to
which, e.g., for $\Sigma _{1}$

\[
\frac{1}{R}=\overline{\stackunder{j\rightarrow \infty }{\lim }}\left|
a_{j}\right| ^{1/j}, 
\]
where the bar denotes the largest limit for the sequence $\left\{ \left|
a_{j}\right| ^{1/j}\right\} .$

In Fig. 2 the obtained solution for $\hat{f}(k)$ is shown in a linear (at
the top) and in a semi - logarithmic (in the bottom) scales, $\alpha =1.7$.
One can see an oscillatory character of the solution at large $k$; this
property can not be visualized directly from the power series expansion.
Therefore, we also get a large $k-$asymptotic of the solution to Eq. (1.9).
The derivation of it is presented in Appendix A in detail. There we
demonstrate that (i) the asymptotics has an oscillatory character and an
exponentially decreasing amplitude, and (ii) the period of oscillations
coincide with high accuracy with that obtained numerically from power series
expansion, see Fig.2. Thus, with the help of Appendix A we have an
independent evidence of the correctness of our procedure and of the solution
presented in this Section above.

Another evidence of the correctness of the obtained results stems from the
comparison of the general expressions (4.1) - (4.5) with those for the two
particular cases $\alpha =2$ and 1, presented in Sect.. 2, see Eqs. (2.4),
(2.5), and in Sect.. 3, see Eqs. (3.2), (3.3), respectively. Indeed, let us
consider Brownian oscillator at first. By comparing expansion (4.1) at = 2
with that given by Eq. (2.4) one can see that the odd coefficients $c_{2j}$
and $c_{2j+2}$ given by Eq. (2.5) have the same recurrent relation as the
coefficients $a_j$ and $a_{j+1}$ given by Eq. (4.4) have, whereas the even
coefficients $c_{2j+1}$ and $c_{2j+3}$ given by Eq. (2.5) have the same
recurrent relation as the coefficients $b_j$ and $b_{j+1}$ given by Eq.
(4.5) have. Now, let us compare general expressions with those for the
Cauchy oscillator. By comparing expansion (4.1) at $\alpha =1$ with that
given by Eq. (3.2) one can see that the coefficients $c_{3j}$ and $c_{3j+3}$
in Eq. (3.3) obey recurrent relation between the coefficients $a_j$ and $%
a_{j+1}$, see Eq. (4.4), whereas the coefficients $c_{3j-1}$ and $c_{3j+2}$
obey recurrent relation for the coefficients $b_j$ and $b_{j+1}$, see Eq.
(4.5). At last, $c_{3j+1}=0,\,\,j\geq 1$, which is also in agreement with
general expansion, if we set $\alpha =1$ in Eq. (4.1). Thus, we may conclude
that the general expansion (4.1) - (4.5) is in agreement with the particular
cases of Brownian and Cauchy oscillators. However, we have to take in mind
that for the Brownian and Cauchy cases we are able to evaluate coefficient $%
a $ analytically. On the contrary, for an arbitrary we have to estimate this
coefficient numerically with the use of Eq. (4.7). Of course, the results of
numerical and analytical estimates coincide for $\alpha =1$ and 2.

Now we consider two important properties which have been already discussed
for the particular case of the Cauchy oscillator in Sect.. 3, namely, power
law tails and bimodality. Consider power law tails at $x\rightarrow \pm
\infty $ at first. These asymptotics are determined by the first non -
analytical term in the power series expansion (4.1), that is, the term $%
a_{1}|k|^{\alpha +2}$. By making an inverse Fourier transform of this term,
we get

\begin{equation}
f(x)\approx a_1\int\limits_{-\infty }^\infty \frac{dk}{2\pi }\left| k\right|
^{\alpha +2}\exp (-ikx)\;,\quad x\rightarrow \pm \infty ,  \tag{4.8}
\end{equation}

where $a_{1} =\left[ \alpha (\alpha +1)(\alpha +2)\right] ^{-1} $ . This
integral is calculated by passing to the complex plane, the detailed
derivation is presented in Appendix B. The result is

\begin{equation}
f(x)\approx \frac{\sin (\pi \alpha /2)\Gamma (\alpha )}{\pi \left| x\right|
^{\alpha +3}}\quad ,x\rightarrow \pm \infty .  \tag{4.9}
\end{equation}

It follows from Eq. (4.8) that the equilibrium PDF has a power law tail, $%
f(x)\propto \left| x\right| ^{-(\alpha +3)}$ , and, thus the integer moments
of the order greater than 3 diverge. This behavior is strikingly different
from that of a non - linear Brownian oscillator. The ``long tails'' can be
explained qualitatively, if we turn to the Langevin description of the
L\'{e}vy oscillator, The Langevin approach relevant to the FSESE [13]
implies that the non - linear overdamped oscillator is influenced by ''white
L\'{e}vy noise'' $\xi (t)$, whose PDF behaves as $|\xi $$|^{-1-\alpha }$%
\thinspace at $|\xi $$|\rightarrow \infty $ . These ''long tails'' imply
that the large absolute values of the noise occur frequently, which, in
turn, lead to large increments of the coordinate. However, it is also clear
that the PDF of the coordinate $x$ must fall off more rapidly at $%
x\rightarrow \infty $ than the PDF of the noise $\xi $, because of the
presence of the potential well, which prevents $x$ from ''escaping'' rather
far from the origin.

In Fig. 3 equilibrium PDF is shown by solid lines in a linear (at the top)
and semi - logarithmic (at the bottom) scales. The PDF is obtained by an
inverse Fourier transform of the characteristic function shown in Fig. 2, $%
\alpha =1.5$. The dashed lines indicate asymptotic (4.9). One can see,
especially from the semi - logarithmic plot, that the asymptotics is a good
approximation beginning from $k$ equal nearly 2. In this figure the second
important property, namely, bimodality is clearly seen in a linear scale. In
Fig. 4 the profiles of equilibrium PDFs (obtained by an inverse Fourier
transform of corresponding power series in $k$ -space) are shown for the
different L\'{e}vy indexes from $\alpha =1$ at the top of the figure till $%
\alpha =2$ at the bottom. It is seen that the bimodality is most strongly
expressed for $\alpha =1$. With the L\'{e}vy index increasing the bimodal
profile ``smooths'' and, finally, it turns to a unimodal one at $\alpha =2$,
that is, for the Boltzmann distribution.

We also perform numerical simulation based on numerical solution of the
Langevin equation

\begin{equation}
\frac{dx}{dt}=-x^{3}+\xi ,  \tag{4.10}
\end{equation}
where $\xi (t)\,$is a white Gaussian noise or a white L\'{e}vy noise, whose
generator is described in detail in our previous papers devoted to the
studies of self - affine properties of ordinary and fractional L\'{e}vy
motions \cite{Chechkin1},\cite{Chechkin2}. A discussion on equivalency of
the description of a stochastic system with the help of the Langevin
equation (21) and fractional kinetic equation (1) is presented in Ref. \cite
{Chechkin}; for the Brownian motion this problem is discussed, e.g., in Ref. 
\cite{Chandrasekhar}, in detail.

In Fig.5 the results of numerical modelling are presented for the Brownian
oscillator, $\alpha =2$ (above) and the L\'{e}vy oscillator, $\alpha =1.1$
(below). At the left the trajectories $x(N)$ are presented , where $%
N=t/\Delta t$ is the number of time steps in numerical modeling, $t$ is the
length of a single step, $\Delta t<<1$. The result of numerical solution of
the Langevin equation must not depend on $\Delta t$; for the Brownian
oscillator this requirement is fulfilled at $\Delta t\leq 10^{-2}$, whereas
for the L\'{e}vy oscillator $\Delta t\leq 10^{-3}$. We also studied the time
- dependence of the second moments and fix when the moments become constant,
thus indicating equilibrium state. From the left figures a clear difference
between trajectories of Brownian (above) and L\'{e}vy (below) oscillators is
seen: there are large ``jumps'' on the figure below, which are due to
existence of large ``pushes'' from an external L\'{e}vy noise $\xi \,\,$or,
equivalently, due to power law asymptotic of the PDF of the L\'{e}vy noise $%
\xi $. Now we turn to the right figures. The designations are as follows.
Thick solid line 1 and dotted line 2 show the potential well and its
curvature, respectively, in conventional units. Thin solid line 3 shows the
Boltzmann distribution (2.2). The black points depict the PDFs obtained in
numerical simulations by statistical averaging over 50 trajectories each of
20.000 steps. Finally, the power law asymptotic (4.9) are depicted by a thin
solid line 4. It is seen from the figure above that the PDF obtained in
numerical simulations agrees quantitatively with the Boltzmann PDF, whereas
the figure below demonstrates drastic difference between the Boltzmann PDF
and numerical PDF. The latter has long power asymptotic, which start far
away from the maximum of the potential well curvature, and whose exponent is
close to that obtained theoretically. This conclusion is confirmed by the
results of numerical modeling for the L\'{e}vy oscillators with different
L\'{e}vy indexes, see Fig. 6. In this figure the black points depict the
values $\gamma \,\,$of the exponents of the asymptotic $|x|^{-\gamma }$- of
the equilibrium PDFs for $\alpha =1.1,1.3,1.5,1.7$ and 1.9. The values $%
\gamma \,\,$of are estimated as a tangent of a slope angle of a rectilinear
asymptotic in a double logarithmic scale. The statistical averaging is over
50 trajectories, each consisting of 20.000 time steps, $\Delta t=10^{-3}$.
Dotted line depicts theoretical dependence $\gamma =\alpha +3$ of the
asymptotic exponent versus $\alpha $. The results of numerical simulations
coincide with analytical estimates within the error limits.

In Fig. 7 the results of numerical modeling are presented in more detail for
small values of $x$, when power law asymptotic is inadequate. Similarly to
Fig. 5, thick solid line 1 and dotted line 2 show the potential well and its
curvature, respectively. Solid line 3 indicates PDF obtained by an inverse
Fourier transform of the power series in $k-$space. The black points depict
PDF obtained in numerical simulations by statistical averaging over 20
trajectories , each of $10^{-5}$ steps. The L\'{e}vy index is 1.2. It is
clearly seen that both PDFs are in qualitative agreement and the both are
bimodal. Therefore, from Figs. 5 - 7 we may conclude that the results of
simulations based on numerical solution of the Langevin equation confirm
qualitatively and even quantitatively the results based on analytical
solution of Eq. (1.9) for the characteristic function.

\section{Discussion and Results}

In this paper we study, both analytically and numerically, the properties of
equilibrium PDF of a non - linear ($x^{4}$) L\'{e}vy oscillator, that is,
the oscillator which is subjected to a white L\'{e}vy noise obeying a
L\'{e}vy stable probability law. We restrict ourselves to the case of the
L\'{e}vy indexes such that $1\leq \alpha \leq 2$. It is known that the
L\'{e}vy stable distributions (as the Gaussian one, which corresponds to $%
\alpha \,\,$equal 2) appear in problems, whose result is determined by the
sum of a great number of independent identical factors. Since the Brownian
oscillator is subjected to a white Gaussian noise, the L\'{e}vy oscillator
is a natural generalization of a Brownian one.

For the analytical studies the starting equation is the so - called
fractional symmetric Einstein - Smoluchowski equation, which contains
fractional symmetric space derivative and is a natural generalization of the
kinetic Einstein - Smoluchowski equation used in the theory of Brownian
motion.

The main results are as follows:

1. We get analytically the characteristic function of the equilibrium PDF in
the form of a power series. Its inverse Fourier transform, realized
numerically, allows us to obtain the PDF. The two main distinctive features
of the equilibrium PDFs for non - linear L\'{e}vy oscillator with $\alpha
\neq 2$ are (i) the power law asymptotic at large x, and (ii) bimodality.
Both features imply that the PDFs for non - linear L\'{e}vy oscillators are
strikingly different from Boltzmann distribution, which is equilibrium PDF
for the Brownian oscillator.

2. The power law asymptotics is determined by the first non - analytical
term in a power series expansion of the characteristic function. We find
that the asymptotics behaves as $|x|^{-\left( \alpha +3\right) }$.

4. The bimodal profile of the equilibrium PDF is clearly visualized after an
inverse Fourier transform of the power series expansion for the
characteristic function. Two maxima are most strongly expressed for $\alpha
=1$ (Cauchy oscillator). With the L\'{e}vy index increasing the profile
``smooths'' and at $\alpha =2$ (Brownian oscillator) turns to profile with a
single maximum.

5. We make three independent verifications of the solution for the
characteristic function.

First: we find that the power series expansion for the characteristic
function of the Brownian oscillator ($\alpha =2$) is in agreement with the
solution for $1\leq \alpha \leq 2$, if we set $\alpha =2$ in the latter.

Second: the particular case of the Cauchy oscillator ( $\alpha =1$) admits
complete analytical study, which allows us also to check the agreement
between power series and to demonstrate power law asymptotics and bimodality
of the PDF.

Third: we obtain the solution of the equation for the characteristic
function at large values of the argument $k$ and show that the periods of
oscillations of this asymptotic solution in $k-$space coincide with high
accuracy with those obtained from the power series.

Therefore, all three particular verifications testify to correctness of our
approach in the general case, $1\leq \alpha \leq 2$.

6. We also perform numerical simulation based on numerical solution of the
Langevin equation for a non - linear oscillator subjected to a white
Gaussian noise. The equilibrium PDFs obtained in simulation show
quantitative agreement with PDFs obtained analytically.

At the end we note that the general case of a non - linear L\'{e}vy
oscillator ($x^{2n},\,\,n=1,2,..$.) can be treated in a similar way. In
particular, it can be shown that in this case the power law asymptotics have
an exponent $-$ $(\alpha +2n-1)$.

\strut

{\Large ACKNOWLEDGMENTS}

Information support by the Project INTAS 98 - 01 is acknowledged.

\strut

{\Large APPENDIX A. Asymptotics of equilibrium characteristic function at
large }$k$

We start from Eq. (1.9) for $k>0$. Using the transformation \cite{Kamke}

\begin{equation}
\eta (\xi )=k^{b/3}\hat{f}(k),\quad \xi =ck^{1+b/3},  \tag{A.1}
\end{equation}
where $b=\alpha -1$, and $c$ is an arbitrary positive (for definiteness)
parameter, we get

\begin{equation}
\xi ^{3}\eta ^{^{\prime \prime \prime }}+(1-\nu ^{2})\xi \eta ^{^{\prime
}}+\left( \nu ^{2}-1-\frac{a\nu ^{3}}{c^{3}}\xi ^{3}\right) \eta
=0\,,\,\,\,\,\,\,\nu =\frac{3}{\alpha +2}.  \tag{A.2}
\end{equation}
At large Eq. (A.2) reduces to

\[
\eta ^{^{\prime \prime \prime }}-\left( \frac{\nu }{c}\right) ^{3}\eta =0, 
\]
whose solution, which tends to zero at large is

\begin{equation}
\eta (\xi )=C_{1}\exp \left[ \frac{\nu }{c}\xi \left( -\frac{1}{2}+\frac{i%
\sqrt{3}}{2}\right) \right] +C_{2}\exp \left[ \frac{\nu }{c}\xi \left( -%
\frac{1}{2}-\frac{i\sqrt{3}}{2}\right) \right] .  \tag{A.3}
\end{equation}
where $C_{1}$ and $C_{2}$ are arbitrary constants. Returning to $\hat{f}(k)$
, we get at large $k$

\begin{equation}
\hat{f}(k)=Ck^{-(\alpha -1)/3}\exp \left[ -\frac 3{2(\alpha +2)}k^{(\alpha
+2)/3}\right] \times \cos \left[ \frac{3\sqrt{3}}{2(\alpha +2)}k^{(\alpha
+2)/3}-\theta \right] .  \tag{A.4}
\end{equation}

It depends on two unknown constants, $C$ and $\theta $, since we use the
condition at infinity only. We can check this formula for two particular
cases, $\alpha =1$ and 2.

1. Cauchy oscillator, $\alpha =1$.

By comparing Eqs. (A.4) and (3.1) we get $C=2/\sqrt{3} \;,\quad \theta =\pi
/6$ .

2. Brownian oscillator, $\alpha =2$.

We explore asymptotics of the following integral at large $k$, which can be
obtained with the help of the saddle - point method, see, e.g., \cite{Olver}%
: 
\[
\int\limits_{-\infty }^{\infty }dx\exp (-x^{2m}+ikx)=\frac{2\sqrt{\pi }}{%
\sqrt{m(2m-1)}}(2m)^{(m-1)/(2m-1)}k^{-(m-1)/(2m-1)}\times 
\]
\begin{equation}
\exp \left( k^{2m/(2m-1)}c_{m}\cos \frac{\pi m}{2m-1}\right) \cos \left(
k^{2m/(2m-1)}c_{m}\sin \frac{\pi m}{2m-1}-\frac{\pi }{2}\frac{m-1}{2m-1}%
\right) \quad ,  \tag{A.5}
\end{equation}
where $m=1,2,...,$

\[
c_{m}=\frac{2m-1}{2m}\left( 2m\right) ^{-1/(2m-1)}. 
\]
With the help of Eq. (A.5) we get, also using Eqs. (2.2), (2.3),

\[
\hat{f}(k)=\int\limits_{-\infty }^\infty dxf(x)\exp (ikx)=\frac{\sqrt{2}}{%
\Gamma (1/4)}\int\limits_{-\infty }^\infty dx\exp \left( -\frac{x^4}%
4+ikx\right) \approx 
\]

\begin{equation}
\approx \frac{2\sqrt{\pi }}{\sqrt{3}\Gamma (1/4)}\left| k\right| ^{-1/3}\exp
\left( -\frac{3}{8}\left| k\right| ^{4/3}\right) \cos \left( \frac{3\sqrt{3}%
}{8}\left| k\right| ^{4/3}-\frac{\pi }{6}\right)  \tag{A.6}
\end{equation}
at $|k|>>1$. By comparing Eqs. (A.6) and (A.4) we get $C=2\sqrt{\pi }/(\sqrt{%
3}\Gamma (1/4))\;,\quad \theta =\pi /6.$

We also make a comparison between the period of the asymptotics (A.4) and
the period of oscillations of the solution (4.1) - (4.7). Firstly, we are
able to get with high accuracy the values $k_j,\,j=1,2,3,4,5$, at which the
solution given by Eqs. (4.1) - (4.7) is equal zero, see also Fig. 2. Then,
we insert $k_j$ into the cosine in Eq. (A.4) and estimate

\begin{equation}
1-\frac{3\sqrt{3}}{2(\alpha +2)}\frac{k_{j+1}^{(\alpha +2)/3}-k_j^{(\alpha
+2)/3}}\pi =\delta _{j+1,j}\,.  \tag{A.7}
\end{equation}

It is seen, that $\delta _{j+1,j}$ can serve as a measure of difference
between zeros of $\hat{f}(k)$ estimated from Eqs. (4.1) - (4.7) and those
estimated from Eq. (A.4). For example, for the characteristic function with $%
\alpha =1.7$, which is shown in Fig. 2, we get $\delta _{2,1}=56\cdot
10^{-4},\,\delta _{3,2}=17\cdot 10^{-4},\,\,\delta _{4,3}=8\cdot
10^{-4},\,\delta _{5,4}=5\cdot 10^{-4}.$ These results demonstrate the
smallness of $\delta _{j+1,j}$ and serve as one more confirmation of the
correctness of our approach.

\strut

{\Large APPENDIX B. Power law asymptotics of equilibrium PDF}

In this Appendix we evaluate the main value of the integral 
\begin{equation}
I=\int\limits_{-\infty }^{\infty }\frac{dk}{2\pi }\left| k\right| ^{\alpha
+2}\exp (-ikx),  \tag{B.1}
\end{equation}
which gives the asymptotics of the equilibrium PDF, see Eq. (4.8). We
present (B1) as

\begin{equation}
I=2\limfunc{Re}I_{1},  \tag{B.2}
\end{equation}
where

\begin{equation}
I_1=\int\limits_0^\infty \frac{dk}{2\pi }k^{\alpha +2}\exp (-ikx).  \tag{B.3}
\end{equation}

In order to evaluate $I_{1}$, we pass to the complex plane. Since the
integral over the closed contour shown in Fig. 8 is equal zero, we get 
\begin{eqnarray}
I_{1} &=&-\int\limits_{-i\infty }^{0}\frac{dk}{2\pi }k^{\alpha +2}\exp
(-ikx)=i\exp (-i\alpha \pi /2)\int\limits_{0}^{\infty }\frac{dk}{2\pi }%
k^{\alpha +2}\exp (-kx)=  \nonumber \\
= &&\frac{i\exp (-i\alpha \pi /2)}{2\pi x^{\alpha +3}}\Gamma (\alpha +3) 
\tag{B.4}
\end{eqnarray}
and 
\begin{equation}
I=\frac{\sin \left( \frac{\pi \alpha }{2}\right) \Gamma (\alpha +3)}{\pi
x^{\alpha +3}}.  \tag{B.5}
\end{equation}
With the use of Eqs. (4.8) and (B.5) we get power law asymptotics of the
equilibrium PDF, see Eq. (4.9).

\end{document}